\documentclass[12pt]{amsart}

\def\Z{\mathbb Z}
\def\C{{\mathbb C}}

\def\acts{\triangleright}
\def\CA{{\mathcal A}}
\def\CH{{\mathcal H}}

\def\ei{e^{2\pi i \theta}}
\def\emi{e^{-2\pi i \theta}}
\def\ket#1{| #1 \rangle}

\theoremstyle{remark}
\newtheorem{rem}[]{Remark}
\def\ts{\otimes}

\def\oq{\frac{1}{q}}
\def\oqd{\tfrac{1}{q^2}}

\def\endremark{\vrule height 0.5em depth 0.2em width 0.5em}
\title{More Noncommutative 4-Spheres}
\author{Andrzej Sitarz}
\email{Andrzej.Sitarz@th.u-psud.fr}
\thanks{${}^\dagger$\ Supported by Marie Curie Fellowship.}
\address{Laboratoire de Physique Theorique, Universit\'e Paris-Sud, 
Bat. 210, 91405  ORSAY Cedex, France}
\begin{document}
\begin{abstract}
New examples of noncommutative 4-spheres are introduced. 
\end{abstract}
\maketitle
\section{Introduction}

Recently some new examples of noncommutative algebras, which 
are deformations of spheres (3 nad 4-dimensional) were presented 
\cite{CL,DLM}. The significant part of the construction was the 
deformation of the 3-sphere, then the 4-sphere in both cases 
is obtained by a commutative suspension.  

The deformation \cite{CL}, which, on the algebraic level is a twist 
of the algebra of functions relative to the action of two $U(1)$ 
groups, preserves much of the geometry of the sphere. 
In particular, the spectral triple is also deformed with the 
Dirac operator unchanged (similarly as for the noncommutative 
torus), which explains the name of an {\em isospectral 
deformation} \cite{CL}.

In this Letter we propose another family of noncommutative 
4-spheres constructed using the method of "twisting". Let us 
stress, however, that such deformed algebras will probably 
not be 'noncommutative spin manifolds' as described by spectral 
triples (see \cite{Con,Con2} for details) and their geometry
seems to be much different from the commutative four-sphere.

\section{Noncommutative 2-sphere}
Let us remind here the construction of the Podle\'s quantum 
sphere\footnote{We have chosen here a particular quantum sphere.}
\cite{Pod}. Let $q$ be a real number $0<q \leq 1$ and $S_q$ 
be an algebra generated by operators $a$, $a^\ast$ and 
$b=b^\ast$, which satisfy the following relations:
\begin{equation}
\begin{array}{lcl}
ba = q ab,   & \;\;\;\;\; & a^\ast b = q ba^\ast, \\
a^\ast a + b^2 =1, &\;\;\;\;\; & a a^\ast + \oqd b^2 = 1.
\end{array}
\end{equation}
This algebra admits the action of the $U_q(su(2))$ quantum group
and in the $q=1$ limit corresponds to the continuous functions on
the two-dimensional sphere. 

\section{Noncommutative 4-spheres}

Let us consider an algebra generated by $\alpha, \alpha^\ast, 
\beta=\beta^\ast$ and $U, U^\ast$ with the following contraints:
\begin{equation}
\begin{array}{lll}
\beta \alpha = q \alpha \beta, & \;\;\;\;\; & 
\alpha^\ast \beta =  q \beta  \alpha^\ast,\\
\alpha^\ast \alpha + \beta ^2 + U^\ast U=1, & \;\;\;\;\; &
\alpha \alpha^\ast + \oqd \beta ^2 + U^\ast U= 1, \\
U \alpha = \ei \alpha U, & \;\;\;\;\; & 
U \alpha^\ast = \emi \alpha^\ast U, \\
U \beta = \beta U, & \;\;\;\;\; & U^\ast U = U U^\ast. 
\end{array}
\end{equation}

This defines a family of algebras $S^4_{q,\theta}$. By taking the
supremum of operator norm over all admissible representations,
taking out the ideal of zero-norm elements and completing
the quotient we obtain a $C^\ast$ algebra structure.

Let us make here some remarks.

\begin{rem}
The center of the algebra $S^4_{q,\theta}$ (for $0<q<1$ and irrational $\theta$)
is generated by a non-negative operator $UU^\ast$. The space of characters, 
as one can easily verify, contains all following maps $\chi$:
\begin{equation}
\begin{aligned}
\chi(\alpha) \in S^1, \chi(\beta)=0, \chi(U)=0,  \\
\text{or} \\
\chi(\alpha) =0,  \chi(\beta)=0, \chi(U) \in S^1,
\end{aligned}
\end{equation}
so that the 'classical' space underlying the noncommutative construction is
a union of two disjoint circles. \endremark
\end{rem}

\begin{rem}
In the $\theta=0$ limit $S^4_{q,0}$ is the complex (2-dimensional) 
suspension of the Podle\'s sphere. Note that this is not equivalent to the 
real (1-dimensional) suspension of $SU_q(2)$ constructed in \cite{DLM}.\endremark
\end{rem}

\begin{rem}
A map $\rho: S^4_{q,\theta} \to S^3_\theta$ defined by $\rho(\beta)=0$
and $\rho(\alpha)=\widetilde{\alpha}$, $\rho(U) = \widetilde{\beta}$ (for 
definitions of the generators of $S^3_\theta$ see \cite{CL}) 
is a star algebra morphism.\endremark
\end{rem} 

\begin{rem}
In the $q=1$ limit $S^4_{1,\theta}$ is exactly the sphere $S^4_\theta$ 
of \cite{CL} (obtained through the suspension od $S^3_\theta$). \endremark
\end{rem}

\begin{rem}
The sphere $S^4_{q,\theta}$ arises as a subalgebra of the 
crossproduct algebra from nontrivial action of $\Z$ 
on the one-dimensional suspension of the Podle\'s sphere.
\begin{equation}
\begin{aligned}
n \acts a = \ei a, \\
n \acts a^\ast = \emi a^\ast, \\
n \acts b = b, \\
n \acts t = t,
\end{aligned}
\end{equation}
where $t$ is a central selfadjoint {\em suspension} generator. The
crossproduct algebra of $\C\Z$ is generated by  $a,a^*,b,t$ and
a unitary generator $V$ of $\C\Z$. It is easy to observe that the 
subalgebra generated by $a,a^*, b$ and $Vt=tV$ is isomorphic to 
$S^4_{q,\theta}$. \endremark
\end{rem}

As one can see, the family of 4-spheres we have just introduced,
shares both the properties of the {\em $q$-deformation} as well as 
that of the {\em twisted} deformation.  Notice that for $\theta=0$ 
(as well as for the suspension of $SU_q(2)$ in \cite{DLM}) $U_q(su(2))$ is
still a symmetry group. This does not hold, however, for (irrational) 
$\theta \not=0$.

\section{Representation of $S^4_{q,\theta}$}

To obtain the Hilbert space representation of $S^4_{q,\theta}$ we proceed as
follows. First, let $\CH_q$ be the Hilbert space representation of the Podle\'s 
sphere and $\rho$ the corresponding representation map 
(see \cite{Pod} for details on the representations of $S^2_q$). 
Then for every $0 \leq \phi < 2\pi$ we construct the following representation 
$\rho_\phi$ on $\CH_q \otimes l^2(\Z)$:
\begin{equation}
\begin{aligned}
\rho_\phi(\alpha) \left( v \ts \ket{n} \right)= (\cos \phi) e^{-2 \pi i n \theta} \left(  \rho(\alpha)(v) \ts \ket{n}\right), \\
\rho_\phi(\beta) \left( v \ts \ket{n}\right) = (\cos \phi) \left(  \rho(\beta)(v) \ts \ket{n}\right), \\
\rho_\phi(U) \left( v \ts \ket{n}\right) = (\sin \phi) \left(  v \ts \ket{n+1}\right).
\end{aligned}
\end{equation}

Given this example, let us finish this section with a remark that 
using various representations of $S^2_q$ (which may 
correspond to nontrivial projective modules over the 
quantum sphere) one might obtain many interesting 
models for field theory.   

\section{The projector.}

A question, which is interesting in itself is the construction of 
{\em noncommutative vector bundles} over our noncommutative space. 
This is expressed algebraically by construction of finitely generated 
projective modules over an algebra $\CA$, defined by idempotents in the 
algebra $M_n(\CA)$.  We shall provide here an example of an idempotent 
in $M_4(\CA)$ and calculate its first Chern class in order to make 
contact with the construction of \cite{CL,DLM}. 

Let us take:
\begin{equation}
e = \frac{1}{2} \left( \begin{array}{cccc} 
1+ \beta &0  &U & \alpha^\ast  \\
0 & 1+\oq  \beta & \alpha &  -\ei U^\ast \\
U^* & \alpha^\ast & 1 - \beta & 0  \\
\alpha & -\emi U & 0 & 1 - \oq  \beta 
\end{array} \right),
\end{equation}
We may easily calculate its Chern-Connes character in the reduced $(b,B)$ 
standard complex of the algebra $S^4_{q,\theta}$ using the Dennis trace 
map \cite{Hus}. Clearly, $ch_0(e) = 0$ and:
\begin{equation}
\begin{aligned}
ch_1(e) \sim  &\left(1-\oq \right)  \left( \beta \ts ( U \ts U^* - U^* \ts U) +  \phantom{\oq} \right. \\
& \left.  \phantom{\oq} +  u \ts ( U^* \ts \beta - \beta \ts U^*) +  U^* \ts (  \beta \ts U- U \ts \beta) \right). 
\label{ch1}
\end{aligned}
\end{equation}

Notice that here, $ch_1(e)$ does not depend on $\theta$ and vanishes for $q=1$, 
exactly as in \cite{DLM}. In fact, the formula (\ref{ch1}) is almost  the same as for
the suspension of $SU_q(2)$. In the commutative limit $q=1$ and $\theta=0$ 
we recover the projector of the instanton bundle. Note that $ch_2(e)$ does 
not vanish and contains 222 terms.

Of course, this is not the only projector one can find, in particular,
using the monopole projector for the quantum sphere  we define,
similarly as in \cite{DLM}:

\begin{equation}
e' = \frac{1}{2} \left( \begin{array}{cccc} 
1+ x & 0 & \beta & \alpha^* \\
0 & 1+ x  & \alpha & -\oq  \beta \\
\beta & \alpha^*  & 1- x  & 0 \\
\alpha & -\oq  \beta & 0 & 1- x \end{array} \right),
\end{equation}

where $x = \sqrt{U^*U}$ is a central element.  However, 
this projector lives effectively on the subalgebra of the 
$S^4_{q,\theta}$, which could be identified as a deformation 
of a 3-sphere and therefore it would correspond to
a trivial bundle over $S^3$  \cite{DL}. In fact, it is easy to verify 
that all its Chern classes vanish.

\section{Conclusions}

In this letter we have shown that there exists a large family of objects, which 
could be called {\em noncommutative 4-spheres}: this includes 
the $\theta$-deformation ("isospectral deformation") defined 
in \cite{CL} and extends the family of objects presented in \cite{DLM}. 

All of them have the same commutative limit and all are deformations, 
however, only in the case of the "isospectral" deformation ($q=1$) 
the lower chern characters of the deformed instanton bundle projector 
vanish.

Such objects should be studied in greater details: the classes and examples
of projective modules over the algebra, their cyclic cohomology  etc.
It would be interesting to verify whether these deformations could be 
understood as {\em noncommutative manifolds} in the sense of 
\cite{Con2}. Let us remark that despite some negative results like dimension 
change in cyclic cohomology \cite{Mas} or no-go theorems for some differential
calculi \cite{Sch} the main problem of understanding such deformations 
as noncommutative manifolds might be algebraic and related with the choice 
of the dense subalgebra of  "$C^\infty$" functions on our deformations (see 
Rieffel's "Question 19" in \cite{Rie}).

\end{document}